\documentclass[12pt]{iopart}

\usepackage{graphicx}

\usepackage{url}

\begin{document}

\title[Preprint - ArXiv]{The Lorentz factor in a reverse coordinate system}

\author{D Roldán$^{1}$\footnote{diego.roldan@ucuenca.edu.ec}, R Sempertegui$^{1}$\footnote{rodrigo.sempertegui@ucuenca.edu.ec}, F Roldán$^{1}$\footnote{francisco.roldan@ucuenca.edu.ec}\\
\small{$^{1}$University of Cuenca}\\
}

\vspace{10pt}
\begin{indented}
\item[]March 2022
\end{indented}

\begin{abstract}
In the present study, we have derived the Lorentz factor using a coordinate system with antiparallel X-axes. Using a thought experiment, common in relativistic literature, we have used the case of a pulse of light moving along the X-axis. Next, we have argued that, consequently with the isotropy of space, the result must be the same if the trajectory of the light pulse is any angle $\alpha>0^{\circ}$, thus obtaining an alternative transformation factor that generates the same results as the Lorenz factor at any angle of the trajectory of the light pulse, therefore confirming the relativistic isotropy of space in the context of the Minkowski spacetime. Nevertheless, a particularly important novelty is that the derived alternative transformation factor allows the reference frames involved to move at speeds greater than that of light.
\end{abstract}
%
\vspace{2pc}

\section{Introduction}

\begin{figure}
\begin{center}
\includegraphics[width=0.80\textwidth]{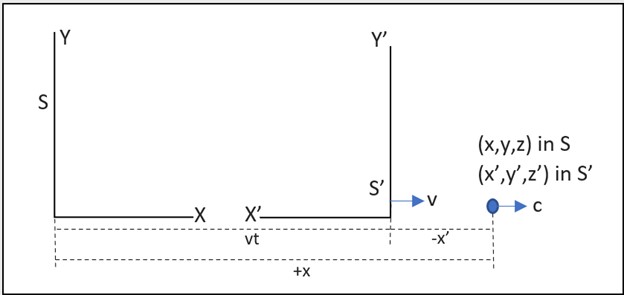}
\end{center}
\caption{Reverse frames. Location of a point from the perspective of two inertial systems with X-axes in opposite directions.} \label{figure1}
\end{figure}

The Lorentz factor is fundamental in relativistic physics. In a usual way, this mathematical expression is derived using two coordinate systems  with parallel axes, a condition that, according to Friedman and Scarr \cite{ref01}, breaks the symmetry in the speed of the reference frames since frame S moves in a negative direction with respect to frame S', which means that the structure of the transformations equations and their inverse transformation are not the same \cite{ref02}. In any case, the selection of any coordinate system does not imply errors in the calculations, but it rather  facilitates its processes and highlights features such as those related to symmetry. Friedman and Scarr \cite{ref01} propose using coordinate systems in which the X-axes are antiparallel, i.e. parallel lines with opposite directions as can be seen in  Figure \ref{figure1}. These coordinates will be called reverse coordinate systems or reverse frames.

In this study, we derive the Lorentz factor between two inertial frames S and S' using reverse frames and mathematizing the two postulates of special relativity (SR): the relativity of motion and the constancy of the speed of electromagnetic waves in a vacuum \cite{ref03}. As a consequence of this derivation, we obtain a transformation factor very similar to that of the Lorentz factor (LF) within the Minkowski spacetime; these factors largely coincide in terms of their results (calculations), but the new factor presents one additional significant characteristic: opens up possibilities for superluminal motion. According to Dingel et al. \cite{ref04}, when different relativistic effects are combined “it is well-known that richer and bizarre SR behaviours are produced.” In our case, we will just show some results of theoretical interest.

In Figure \ref{figure1}, a pulse of light that starts from the origin O of reference frame S with an angle $\alpha=0^{\circ}$ (angle between the path of the light pulse and the X-axis) and which after a time $t$ is at location (x,y,z), i.e. the light pulse moves along the X-axis. This same pulse from reference frame S' has an angle $\alpha=180^{\circ}$, i.e., it moves in the negative direction of the X'-axis. Except for the use of the reverse coordinate system, this scenario is the typical one established to derive the Lorentz transformations (LT).

Since 1905, many methods of derivation of the LT have been investigated that have enriched our understanding. We undertake a derivation based on the two principles of SR, that is, the principle of relativity and the constancy of the speed of light—or more specifically the "inertial invariance of the speed of propagation of electromagnetic radiation in vacuum" \cite{ref03}—considering the special case described in Figure \ref{figure1}. 

\section{Methodology}

We resort to a thought experiment that is common in studies of SR \cite{ref05}. It is important to emphasize that our study is limited to the context of the special theory of relativity within the Minkowski spacetime, to all phenomena without significant influence of gravitation. We assume two inertial reference frames: S considered at rest, and S’ with inertial motion and velocity $v$ along the X-axis.

We assume that when the origin O' of S' coincides with the origin O of S at the instant $t'=t=0$, a pulse of light is emitted from this origin O with an angle $\alpha=0^\circ$, i.e. in the direction of the X-axis positive. After a time $t>0$, S' will have travelled a distance $vt$ from the perspective of S, at the same time, S will have moved $v't'$ from the perspective of S'. This implies that for S’, the angle of the trajectory of the light pulse is  $\alpha'=180^\circ$.

Without considering the SR postulates, that is, from a Galilean perspective, to derive the transformation equations that allow us to convert the measures of S into those of S', the following equations are initially proposed: 

\begin{eqnarray}
-x'=(x-vt), \label{eq1} \\
x'=(vt-x), \label{eq2} 
\end{eqnarray}

to convert the measures of S’ into those of S

\begin{eqnarray}
x=(v't+(-x')), \label{eq3} \\
x=(v't-x'), \label{eq4} 
\end{eqnarray}

Where the vectors $v=v'$. i.e. frame S sees S' moving away along its positive X-axis and frame S' sees S moving away along its positive X-axis.

Equations (\ref{eq2}) and (\ref{eq4}) represent the Galilean transforms in reverse coordinates. Note that, in this case, the two equations have the same structure highlighting the symmetry between them.   

Now, according to the postulates of SR, it is assumed that the speed of light is constant, and that time is not absolute and can be different between the frames S and S' ($t \neq t'$). On the other hand, by the first postulate of the SR, the relationship must be the same (same transformation factor $\gamma$ from any of the reference frames). Therefore,

\begin{eqnarray}
x'=\gamma(vt-x), \label{eq5} \\ 
x=\gamma(v't'-x'), \label{eq6} 
\end{eqnarray}

The objective is to obtain the values  of the factor $\gamma$ of equations (\ref{eq5}) and (\ref{eq6}) by identifying special \textit{“known”} cases \footnote{ \url{http://web.mst.edu/courses/physics357/Lecture.3.Relativity.Lorentz.Invariance/Lecture3.pdf} } (or events) concordant with the postulates of SR. Each special case is applied to these equations, and subsequently, the values of $\gamma$ is obtained for general use.

\subsection{Derivation of Lorentz factor}

\begin{figure}
\begin{center}
\includegraphics[width=0.80\textwidth]{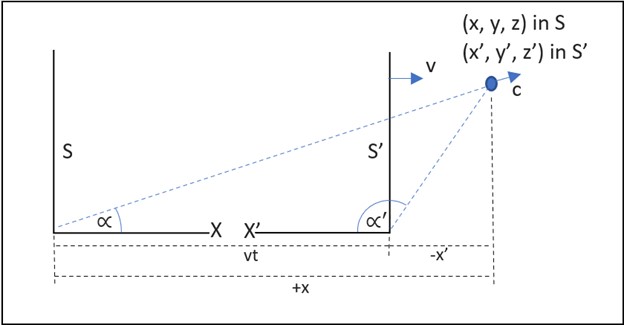}
\end{center}
\caption{Reverse coordinate system. Location of a point from the perspective of two inertial systems with X-axes in opposite directions. With a pulse of light with a trajectory at an angle $\alpha>0^\circ$.} \label{figure2}
\end{figure}

For the derivation of LTs, a light pulse moving along the X-axis is often used as a thought experiment. This strategy is for mathematical convenience, but not mandatory. According to the principle of isotropy of space, we should obtain the same result if we resort to a light pulse with a trajectory in any direction ($\alpha>0^\circ$) as shown in  Figure \ref{figure2}.

According to the description of the thought experiment used and depicted in  Figure \ref{figure2}, from the second postulate,

\begin{eqnarray}
r=ct , \label{eq7} \\ 
r'=c't' . \label{eq8} 
\end{eqnarray}

It should be noted that the path $r>r'$, therefore $ct>c't'$. And since $c$ is a constant, it geometrically suggests that $t>t'$.

Its components on the X-axis are

\begin{eqnarray}
x=cos\alpha\ ct , \label{eq9} \\ 
-x'=cos\alpha\ (-c')t' , \label{eq10} \\ 
x'=cos\alpha'\ c't' . \label{eq11} 
\end{eqnarray}


We isolate from (\ref{eq9}) and (\ref{eq11}) the time variables

\begin{eqnarray}
t=\frac{x}{cos\alpha \ c} \label{eq12} \\
t'=\frac{x'}{cos\alpha' \ c'} \label{eq13} 
\end{eqnarray}

We substitute, respectively, in (\ref{eq5}) and (\ref{eq6})

\begin{eqnarray}
x'=\gamma (\frac{vx}{cos\alpha\ c} -x)=\gamma x(\frac{v}{cos\alpha \ c} - 1), \label{eq14} \\
x=\gamma (\frac{v'x'}{cos\alpha'\ c'}-x')=\gamma x'(\frac{v'}{cos\alpha'\ c'} -1). \label{eq15} 
\end{eqnarray}

We multiply these equations:

\begin{eqnarray}
x'x=\gamma^2 xx' \left(\frac{v}{cos\alpha\ c}-1\right)\left(\frac{v'}{cos\alpha'\ c'}-1\right) \label{eq16}
\end{eqnarray}

And we obtain

\begin{eqnarray}
\gamma=\frac{1}{\sqrt{\left(\frac{v}{cos\alpha\ c}-1\right)\left(\frac{v'}{cos\alpha'\ c'}-1\right)}} \label{eq17} \\
\gamma=\frac{1}{\sqrt{\left(\frac{v}{cos\alpha\ c}-1\right)\left(\frac{v}{cos\alpha'\ c}-1\right)}} \label{eq18}
\end{eqnarray}

and therefore,  

\begin{eqnarray}
\gamma_R=\frac{1}{\sqrt{\left(1-\frac{v}{cos\alpha\ c}\right)\left(1-\frac{v}{cos\alpha'\ c}\right)}} \label{eq19}
\end{eqnarray}

\subsection{A special case where $\alpha=0^\circ$}

This would be the alternative transforming factor (ATF) that apparently differs from the Lorentz factor. However, if we assume the special case in which the light pulse moves along the X-axis of S, it implies that $\alpha=0^\circ$, therefore $cos0^\circ =1$, corresponds in S' to the angle $\alpha'=180^\circ$ and therefore $cos180^\circ =-1$.

\begin{eqnarray}
\gamma=\frac{1}{\sqrt{\left(1-\frac{v}{1\ c}\right)\left(1-\frac{v}{-1\ c}\right)}} \label{eq20}  \\
\gamma=\frac{1}{\sqrt{\left(1-\frac{v}{c}\right)\left(1+\frac{v}{c}\right)}} \label{eq21}  \\
\gamma_L=\frac{1}{\sqrt{1-\frac{v^2}{c^2}}} \label{eq22} 
\end{eqnarray}

Equation (\ref{eq22}) for reverse coordinates is the same Lorenz factor present in the relativistic literature with parallel coordinates.

As we will see later, the transformation factors (\ref{eq19}) and (\ref{eq22}) generate the same values, so it could be assumed that they are two forms of the same equation, however, equation (\ref{eq19}) allows in some cases values of v greater than those of light, which makes an interesting difference.

\subsection{A special case where $\alpha>0^\circ$}

As we explained before, assuming $\alpha=0^\circ$ is arbitrary since, considering the isotropic principle of space, the results must be the same for any $\alpha>0^\circ$, which we will prove next.

A limitation of the ATF (\ref{eq19}) is that we do not yet have a formula for the transformation between the angles $\alpha$ and $\alpha'$. This relationship can be established with the following analysis:

According to Figure \ref{figure1}, in which

\begin{eqnarray}
tan\alpha =\frac{y}{x} . \label{eq23} 
\end{eqnarray}

since there is no relative motion in the direction of the Y-axis, then
\begin{eqnarray}
y=y'=sin\alpha'\ c't'. \label{eq24} 
\end{eqnarray}

According to (\ref{eq6}),
\begin{eqnarray}
x=\gamma_R (v't'-x'). \label{eq25} 
\end{eqnarray}

Substituting (\ref{eq10}) in (\ref{eq25}),
\begin{eqnarray}
x=\gamma_R (v't'-cos\alpha'\ c't'). \label{eq26} 
\end{eqnarray}

Substituting (\ref{eq24}) and (\ref{eq26}) in (\ref{eq23}),
\begin{eqnarray}
tan\alpha=\frac{sin\alpha'\ c't'}{\gamma_R (v't'-cos\alpha'\ c't') }, \label{eq27} \\ 
tan\alpha=\frac{sin\alpha'}{\gamma_R (\frac{v't'}{c't'}-cos\alpha' \frac{c't'}{c't'} ) }, \label{eq28} \\ 
tan\alpha=\frac{sin\alpha'}{\gamma_R (\frac{v'}{c'} - cos\alpha' )} , \label{eq29}  \\
tan\alpha=\frac{sin\alpha'}{\gamma_R (\frac{v}{c} - cos\alpha' )} . \label{eq30} 
\end{eqnarray}

This equation is remarkably similar to the usual functional form in the relativistic literature \cite{ref06, ref07}, with the difference that the LF is replaced by the new ATF  $\gamma_R$.

For comparative purposes, what has been hitherto analyzed is summarized in Table~\ref{math-tab1}.

\begin{table}
\caption{\label{math-tab1}Current Lorentz factor (LF) and alternative transformation factor (ATF).}
\begin{tabular}{ p{4.2cm} p{1cm} p{5.8cm} p{1cm}} 
\br
Current (LF)& & Alternative (ATF)\\
\mr
 $\gamma_L=\frac{1}{\sqrt{\left(1-\frac{v^2}{c^2}\right)}}$   & (\ref{eq22}) &
 $\gamma_R=\frac{1}{\sqrt{\left(1-\frac{v}{cos\alpha\ c}\right)\left(1-\frac{v}{cos\alpha'\ c}\right)}}$ & (\ref{eq19}) \\ [15pt]
& &
$tan\alpha=\frac{sin\alpha'}{\gamma_R \left(\frac{v}{c}-cos\alpha'\right)}$ & (\ref{eq30})  \\ [17pt]
\br
\end{tabular}
\end{table}

The Lorentz factor (\ref{eq22}) provides the same calculations as (\ref{eq19}) and (\ref{eq30}) combined (column ATF, Table~\ref{math-tab1}); as can be seen, the LF does not depend on the angles involved in the ATF. However, as we will discuss below, this does not necessarily lead to different results—at least not always.

Thus, for example, if we consider a velocity $v=0.3c$ and an angle $\alpha'=150^\circ$  in S’, these data must correspond to an angle $\alpha<(180-\alpha')$ by observation of Figure \ref{figure2}.

According to (\ref{eq22}), 

\begin{equation*}   
\gamma_L=\frac{1}{\sqrt{\left(1-\frac{0.3^2 c^2}{c^2}\right)}}=1.0483
\end{equation*}


According to (\ref{eq19}),

\begin{eqnarray}
\gamma_R=\frac{1}{\sqrt{\left(1-\frac{0.3}{cos\alpha}\right)\left(1-\frac{0.3}{cos150^\circ}\right)}}=\frac{0.86181}{\sqrt{\left(1-\frac{0.3}{cos\alpha}\right)}} \label{eq31}
\end{eqnarray}

According to (\ref{eq30}),
\begin{eqnarray}
tan\alpha=\frac{sin150^{\circ}}{\alpha_R(0.3-cos150^{\circ})}=\frac{0.428807}{\gamma_R}     \label{eq32}
\end{eqnarray}

According to (\ref{eq31}) and (\ref{eq32}),

\begin{eqnarray}
\frac{0.86181}{\sqrt{\left(1-\frac{0.3}{cos\alpha}\right)}} =\frac{0.428807}{tan\alpha}     \label{eq33}
\end{eqnarray}

With numerical methods, the angle that satisfies equation (\ref{eq33}) is $\alpha=22.247307^\circ$ and that corresponds to a factor $\gamma_R=1.0483$, that is, the same value as the Lorentz factor $\gamma_L$. We highlight the fact that the relationship of the angles  $\alpha$ and $\alpha'$ agrees with that established by the phenomenon of relativistic aberration within the Minkowski spacetime (i.e., even in the absence of gravity).

In general, once the calculations have been made, the two systems of equations provide the same results; this suggests that LF (\ref{eq22}) can be derived by combining (\ref{eq19}) and (\ref{eq30}); however, the ATF presents some novelties.

\subsection{Faster than the speed of light $v>c$}
\
An important novelty of the ATF, unlike the LF, is that it allows relative speeds $v$ greater than those of light, although it is not new that in the literature we find some studies that in the context of SR, general relativity, or quantum physics allow us to infer speeds faster than that of light \cite{ref08, ref09}, particularly concerning the case of neutrinos that may be tachyons \cite{ref10, ref11}. 

Thus, for example, according to ATF if the speed of the frame S’ is $v=1.2c$ and the angle $\alpha'=150^\circ$

According to (\ref{eq19}),

\begin{eqnarray}
\gamma_R=\frac{1}{\sqrt{\left(1-\frac{1.2}{cos\alpha}\right)\left(1-\frac{1.2}{cos150^\circ}\right)}}=\frac{0.647437}{\sqrt{\left(1-\frac{1.2}{cos\alpha}\right)}} \label{eq34}
\end{eqnarray}

According to (\ref{eq30}),
\begin{eqnarray}
tan\alpha=\frac{sin150^{\circ}}{\alpha_R(1.2-cos150^{\circ})}=\frac{0.2420106}{\gamma_R}     \label{eq35}
\end{eqnarray}

According to (\ref{eq31}) and (\ref{eq32}),

\begin{eqnarray}
\frac{0.647437}{\sqrt{\left(1-\frac{1.2}{cos\alpha}\right)}} =\frac{0.2420106}{tan\alpha}     \label{eq36}
\end{eqnarray}

With numerical methods, the angle that satisfies equation (\ref{eq36}) is $\alpha=210^\circ$ and that corresponds to a factor $\gamma_R=0.419$. This result is interesting as it allows us to speculate a little more about the possibility of hypothetical tachyons \cite{ref12}, partially coinciding with the results of Jin and Lazar \cite{ref09}. 

Currently, the mainstream physics community has not recognized the existence of superluminal phenomena. However, it is not a ruled-out issue and various references can be found in the scientific literature, usually from a theoretical perspective \cite{ref13, ref14, ref15} as in our case.

Ruan \cite{ref16, ref17} proposes in his theory of observational relativity that the maximum speed limit does not correspond to the speed of light $c$ but to the speed of information $\eta$, establishing the observability of the physical world as a principle or axiom, generalizing in this way the Lorentz factor and leaving room "to explore superluminal observational agents" that exceed this observational limit.

\section{Conclusions}

In this study, we have derived the Lorentz factor using a coordinate system with antiparallel X-axes, i.e. parallel, but in opposite directions. Using a thought experiment, frequent in the relativistic literature, we have used the case of a pulse of light that moves along the X-axis, obtaining the Lorenz factor as a transformation factor, confirming the predictability of non-dependence on the chosen coordinate system. In this regard, the AFT can be considered as a novelty way of deriving this factor. 

Next, we have argued that, in consequence of the isotropism of space, the result must be the same whether the angle of the trajectory of the light pulse is along the X-axis, that is $\alpha=0^\circ$, or with any angle $\alpha>0^\circ$, thus obtaining a transformation factor with a more complex mathematical expression that, although it includes in its formulation the angles of the path of the light pulse, nevertheless, provides the same results as the Lorenz factor, regardless of the angle of the trajectory, consequently, confirming the relativistic isotropy of space in the context of the Minkowski spacetime. Nevertheless, a particularly important novelty is that the derived ATF allows the reference frames involved to move at speeds greater than that of light.

In any case, it must be remembered that a large portion of the empirical evidence that supports the predictions of the LT does not contradict the ATF since the solutions of the former coincide in their mathematical estimates with la latter. However, unlike the LF, an important characteristic of the proposed ATF is that it does not prevent the possibility of speeds greater than that of light, offering speculative opportunities for the hypothetical tachyons that are addressed in the literature, as Schwartz \cite{ref10, ref11} does to eventually explain dark energy and dark matter using low energy neutrinos as tachyons. 

Although superluminal motion does not have empirical support, just like the breaking of fundamental symmetries, it is still a possibility in the world of physics, thus, numerous experiments at CERN, are testing for violations of this symmetry \cite{ref18, ref19}, therefore, constitute unfinished business of this science.


\section*{References}

\begin{thebibliography}{1} 

\bibitem{ref01} Y. Friedman, T. Scarr, \textit{Symmetry and Special Relativity}, Symmetry, \textbf{11} 1-15, (2019), \url{https://doi.org/10.3390/sym11101235}

\bibitem{ref02} S. Kichenassamy, \textit{Hot Spots in the Weak Detonation Problem and Special Relativity}, Axioms, \textbf{10}, 4, 1-17, (2021) , \url{https://doi.org/10.3390/axioms10040311}

\bibitem{ref03} W N Mathews, \textit{Seven formulations of the kinematics of special relativity}, American Journal of Physics, \textbf{88} 269-278, (2020), \url{https://doi.org/10.1119/10.0000851}

\bibitem{ref04} B Dingel, A Buenaventura, A Chua, N Libatique , K Murakawa, \textit{Relativistic aberration of light mimicked by microring resonator based optical All-Pass Filter (APF)}, Optik, \textbf{183} 82-91, (2019) , \url{https://doi.org/10.1016/j.ijleo.2018.12.149}

\bibitem{ref05} 	P. Alstein, K. Krijtenburg-Lewerissa, W. R. van Joolingen, \textit{Teaching and learning special relativity theory in secondary and lower undergraduate education: A literature review}, Physical Review. Physics Education Research, \textbf{17}, 2, 1-17, (2021), \url{https://doi.org/10.1103/PhysRevPhysEducRes.17.023101}

\bibitem{ref06} A Jarabo, B Masia, A Velten, C Barsi, R Raskar , D Gutierrez, \textit{Relativistic Effects for Time-Resolved Light Transport}, Computer Graphics Forum, \textbf{34} 8, 1-12, (2015) , \url{https://doi.org/10.1111/cgf.12604}

\bibitem{ref07} K Van Acoleyen , J Van Doorsselaere, \textit{Captain Einstein: A VR experience of relativity}, American Journal of Physics, \textbf{88} 10, pp. 801-813, (2020), \url{https://doi.org/10.1119/10.0001803}

\bibitem{ref08} J M Hill , B J Cox, \textit{Einstein's special relativity beyond the speed of light}, Proceedings of the Royal Society A, \textbf{468}, 4174–4192, (2012) , \url{https://doi.org/10.1098/rspa.2012.0340}

\bibitem{ref09} C Jin, M Lazar, \textit{A note on Lorentz-like transformations and superluminal motion}, Journal of Applied Mathematics and Mechanics, \textbf{95} 7, 690-694, (2015) , \url{https://doi.org/10.1002/zamm.201300162}

\bibitem{ref10} C Schwartz, \textit{Tachyon dynamics for neutrinos?}, International Journal of Modern Physics A, \textbf{33} 10, 1-23, (2018) , \url{http://doi.org/10.1142/S0217751X18500562}

\bibitem{ref11} C Schwartz, \textit{An approach for modeling tachyons with gravitation}, International Journal of Modern Physics A, \textbf{34} 19, 1-18, (2019) , \url{http://doi.org/10.1142/S0217751X19501033}

\bibitem{ref12} R Ehrlich, \textit{Faster-than-light speeds, tachyons, and the possibility of tachyonic neutrinos}, American Journal of Physics, \textbf{71}, 1109-1114, (2003) , \url{http://dx.doi.org/10.1119/1.1590657}

\bibitem{ref13} E. Lentz, \textit{Breaking the warp barrier: hyper-fast solitons in Einstein–Maxwell-plasma theory}, Classical and Quantum Gravity, \textbf{38}, 7, 1-14, (2021) , \url{https://doi.org/10.1088/1361-6382/abe692}

\bibitem{ref14} L. M. Caligiuri, \textit{A new quantum - relativistic model of tachyons}, Journal of Physics: Conference Series, \textbf{1251}, 1, 1-26, (2019), \url{https://doi.org/10.1088/1742-6596/1251/1/012009}

\bibitem{ref15} V. A. Kostelecký, S. Samuel, \textit{Spontaneous breaking of Lorentz symmetry in string theory}, Physical Review D, \textbf{39}, 2, 683-685, (1989) , \url{https://doi.org/10.1103/physrevd.39.683}

\bibitem{ref16} X. Ruan, \textit{Information wave and the t heory of o bservational r elativity}, viXra, 1-38, (2017) 

\bibitem{ref17} X. Ruan, \textit{Observation and Relativity: Why is the Speed of Light Invariant in Einstein's Special Relativity?}, Journal of Beijing Univerity of Technology, \textbf{46}, 1, 82-119, (2020) , \url{https://doi.org/10.11936/ bjutxb2019080005}

\bibitem{ref18} R. Lehnert, \textit{Testing times for space–time symmetry}, 11112016, Online. Available: \url{https://cerncourier.com/a/testing-times-for-space-time-symmetry/}, (2016) 

\bibitem{ref19} N. Russell, \textit{Framing Lorentz symmetry}, 24112004. Online. Available: \url{https://cerncourier.com/a/framing-lorentz-symmetry/}, (2004) 

\end{thebibliography}
\end{document}